\documentclass[a4paper,11pt]{article}
\usepackage{pos}

\usepackage{float}  
\usepackage{physics} 
\usepackage{subcaption}  

\title{Multi-nucleon matrix elements on the lattice with the Feynman-Hellmann theorem}
\ShortTitle{Multi-nucleon matrix elements with the FH theorem}

\author*[a]{N. Humphrey}
\author[a]{K. U. Can}
\author[b]{R. Horsley}
\author[c]{P. E. L. Rakow}
\author[d]{G. Schierholz}
\author[e]{H. St\"{u}ben}
\author[a]{R. D. Young}
\author[a]{J. M. Zanotti}

\affiliation[a]{CSSM, Department of Physics, University of Adelaide, Adelaide SA 5005, Australia}

\affiliation[b]{School of Physics and Astronomy, University of Edinburgh, Edinburgh EH9 3JZ, UK}

\affiliation[c]{Theoretical Physics Division, Department of Mathematical Sciences, University of Liverpool, Liverpool L69 3BX, UK}

\affiliation[d]{Deutsches Elektronen-Synchrotron DESY, Notkestr. 85, 22607 Hamburg, Germany}

\affiliation[e]{Regionales Rechenzentrum, Universit\"{a}t Hamburg, 20146 Hamburg, Germany}

\emailAdd{nabil.humphrey@adelaide.edu.au}

\dedicated{
    \vspace{10 mm}
    ADP-25-17/T1279\\
    DESY-25-070 \\
    LTH 1401
}

\abstract{This work presents the first calculation of the lowest moment of the forward Compton structure function $\mathcal{F}_2$ for a multi-nucleon deuteron-like state using Feynman-Hellmann lattice QCD techniques. Using this result as a prototypical example, we chart a course for the systematic study of multi-nucleon structure by building on techniques developed to optimise the computation of the factorially increasing number of Wick contraction terms required to calculate multi-nucleon matrix elements via lattice QCD.}

\FullConference{The XVIth Quark Confinement and the Hadron Spectrum Conference (QCHSC24)\\
 19-24 August, 2024\\
 Cairns Convention Centre, Cairns, Queensland, Australia\\}

\begin{document}
\maketitle

\section{Introduction}


This study sits within a larger programme of work that seeks to develop more precise and scalable computational probes into the structure of multi-hadron states using the fundamental QCD degrees of freedom. Previous work \cite{humphrey_2022} has explored computational techniques to reduce the computational burden of the factorially growing number of Wick contraction terms required for the calculation of multi-hadron matrix elements in lattice QCD. Here, we apply the lattice QCD Feynman-Hellmann techniques described in Refs.~\cite{chambers_2017,utku_2020,utku_2022,hannaford_gunn_2024,agadjanov_2016} to study the Compton amplitude of multi-nucleon states, using an $NN({}^3 S_1)$ deuteron-like state as a prototypical example. Through our work, we seek to expose new pathways to study multi-hadron phenomena such as nuclear shadowing (as explored in for example Ref. \cite{melnitchouk_1992}). \par
\begin{figure}[H]
    \centering
    \includegraphics[width=\textwidth]{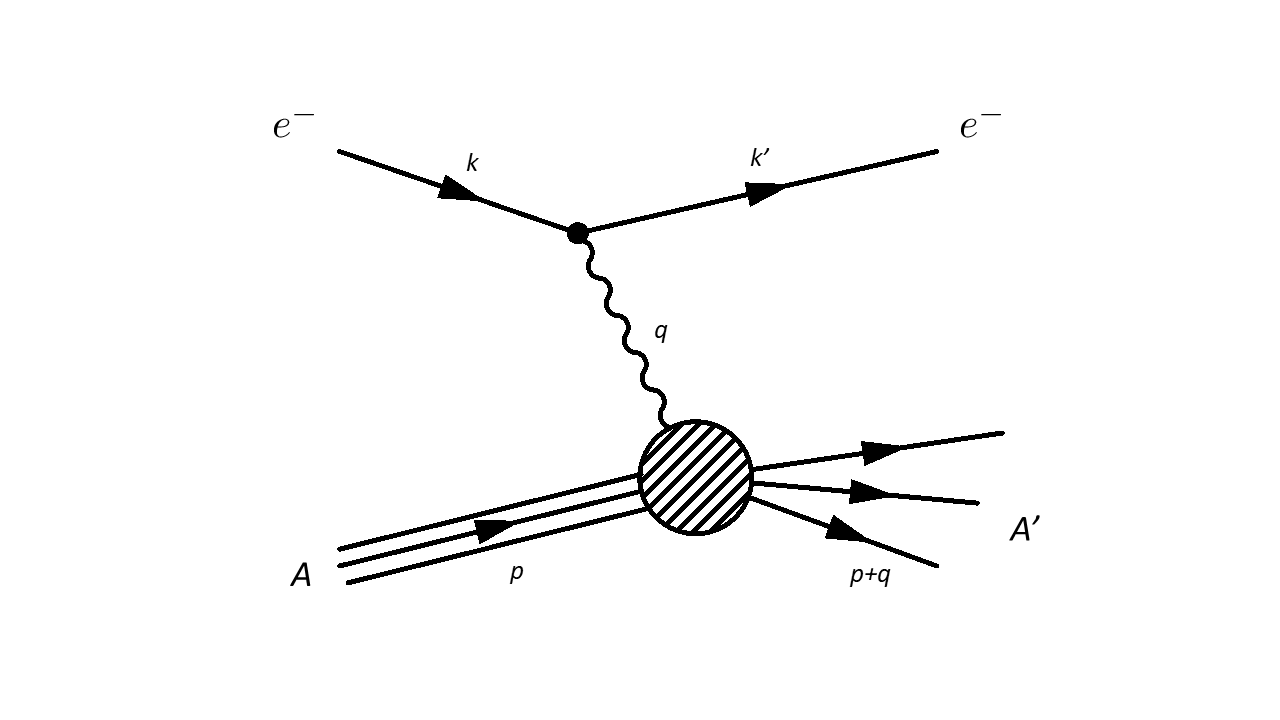}
    \caption{Deep Inelastic Scattering (DIS) diagram for a general multi-hadron target $A$ at momentum $p$.}
    \label{fig:dis_diag}
\end{figure}

Fig.~\ref{fig:dis_diag} depicts the inclusive scattering interaction of a charged lepton $e^-$ from a multi-hadron target $A$, resulting in state $A'$ from a virtual photon transferring momentum $q$. The hadronic target state has momentum $p$, and transitions to a final state with momentum $p+q$.

The strong physics contained within the shaded circle of Fig.~\ref{fig:dis_diag} is encoded by the `hadron tensor' $W_{\mu \nu}$, and the optical theorem related `Compton tensor' $T_{\mu \nu}$, which is conventionally decomposed in the spin-averaged spin-$\frac{1}{2}$ target case in terms of its Lorentz-invariant structure in Minkowski space \cite{hoodbhoy_1988}:
\begin{align}
    T_{\mu \nu}(p,q) &= \frac{i}{2} \int d^z e^{i q \cdot z} \delta_{s s'} \bra{p,s'} \mathcal{T} \left\{ \mathcal{J}_\mu(z) \mathcal{J}_\nu(0) \right\} \ket{p,s} \nonumber \\
    &\eqcolon \left( -g_{\mu \nu} + \frac{q_\mu q_\nu}{q^2} \right) \mathcal{F}_1(\omega, Q^2) + \left( p_\mu - \frac{p \cdot q}{q^2} q_\mu \right) \left( p_\nu - \frac{p \cdot q}{q^2} q_\nu \right) \frac{\mathcal{F}_2(\omega, Q^2)}{p \cdot q},
\end{align}
where $\mathcal{T}$ is the time-ordering operator, $\omega = 2 p \cdot q / Q^2$ is the inverse Bjorken scaling variable, $\mathcal{J}_\mu$ is the electromagnetic current operator, $Q^2 = -q^2$, and $\mathcal{F}_{1,2}(\omega,Q^2)$ are the Compton structure functions.


Similarly, the target polarisation diagonal Compton tensor for a spin-1 target is decomposed as \cite{hoodbhoy_1988}:
\begin{align}
    T_{\mu \nu}^{(\alpha)}(p,q) &= i \int d^z e^{i q \cdot z} \bra{A^{(\alpha)}} \mathcal{T} \left\{ \mathcal{J}_\mu(z) \mathcal{J}_\nu(0) \right\} \ket{A^{(\alpha)}} \nonumber \\
    &\eqcolon \left( -g_{\mu \nu} + \frac{q_\mu q_\nu}{q^2} \right) \mathcal{F}_1(\omega, Q^2) + \left( p_\mu - \frac{p \cdot q}{q^2} q_\mu \right) \left( p_\nu - \frac{p \cdot q}{q^2}q_\nu \right) \frac{\mathcal{F}_2(\omega, Q^2)}{p \cdot q} \nonumber \\
    &\qquad - \frac{b_1(\omega, Q^2)}{(p \cdot q)^2} r_{\mu \nu} + \frac{1}{6} \frac{b_2(\omega, Q^2)}{(p \cdot q)^3 (q \cdot E) (q \cdot E^*)} \left\{ s_{\mu \nu} + t_{\mu \nu} + u_{\mu \nu} \right\} \nonumber \\
    &\qquad + \frac{1}{2} \frac{b_3(\omega, Q^2)}{(p \cdot q)^3 (q \cdot E) (q \cdot E^*)} \left\{ s_{\mu \nu} - u_{\mu \nu} \right\} + \frac{1}{2} \frac{b_4(\omega, Q^2)}{(p \cdot q)^3 (q \cdot E) (q \cdot E^*)} \left\{ s_{\mu \nu} - t_{\mu \nu} \right\} \nonumber \\
    &\qquad + i\frac{g_1(\omega,Q^2)}{p \cdot q} \epsilon_{\mu \nu \lambda \sigma} q^\lambda s^\sigma + i\frac{g_2(\omega,Q^2)}{(p \cdot q)^2} \epsilon_{\mu \nu \lambda \sigma} q^\lambda \left( (p \cdot q) s^\sigma - (s \cdot q) p^\sigma \right) \label{eq:structure_functions},
\end{align}
where $r_{\mu \nu},s_{\mu \nu},t_{\mu \nu},u_{\mu \nu}$ are kinematical tensors that depend in particular on the incoming/outgoing target polarisation direction $(\alpha)$ encoded in 4-vectors $E^\mu$ and $E^{*\mu}$, $s^\sigma$ is an analogous spin vector that is a function of polarisation, and $\mathcal{F}_{1,2}(\omega, Q^2)$, $b_{1,2,3,4}(\omega, Q^2)$, $g_{1,2}(\omega, Q^2)$ are the Compton structure functions. Note that henceforth we will assume that the polarisation direction $(\alpha)$ is in the z-direction, and drop the label from the notation for convenience.

The remainder of this proceedings is organised as follows. §2.1 describes the lattice QCD simulation parameters and choice of interpolating operators. §2.2 provides a short summary of the Feynman-Hellmann Theorem technique for lattice QCD, followed by the specific observables under study in §2.3. §3 presents lattice QCD structure function calculations for a deuteron-like state, and then compares these results against prior experimental and single-nucleon lattice calculations.

\section{Methodology}

\subsection{Simulation Details}

To calculate Compton structure functions, we use a heavier-than-physical quark mass lattice QCD simulation using a single QCDSF gauge ensemble \cite{utku_2020}. The gauge configurations are generated using a stout-smeared non-perturbatively $\mathcal{O}(a)$ improved Wilson action for dynamical quarks and a tree-level Symanzik improved gauge action \cite{utku_2022}. We employ a lattice spacing $a = 0.074(2)$ fm, a lattice volume of $L^3 \times T = 32^3 \times 64$, and a bare coupling parameter of $\beta = 5.5$. We operate at the $SU(3)$ flavour symmetric point, with a pion mass of $\approx 470$ MeV and $m_\pi L = 5.6$. The renormalisation constant was calculated as $Z_V = 0.8611(84)$ \cite{utku_2020}. \par
To calculate observables of multi-nucleon states, we first choose single nucleon interpolating operators at the source:
\begin{align}
    p_\alpha^\pm(x)^\dagger &= \epsilon^{abc} \left[ \left[u^a(x)^\dagger\right]^T (C\gamma_5 P_\pm) d^b(x)^\dagger \right] u^c_\alpha(x)^\dagger, \\
    n_\alpha^\pm(x)^\dagger &= \epsilon^{abc} \left[ \left[d^a(x)^\dagger\right]^T (C\gamma_5 P_\pm) u^b(x)^\dagger \right] d^c_\alpha(x)^\dagger,
\end{align}
with corresponding momentum-projected single nucleon operators at the sink:
\begin{align}
    p_\alpha^\pm(\vec p) &= \sum_{\vec x} e^{-i\vec p \cdot \vec x} \epsilon^{abc} \left[ \left[u^a(x)\right]^T (C\gamma_5 P_\pm) d^b(x) \right] u^c_\alpha(x), \\
    n_\alpha^\pm(\vec p) &= \sum_{\vec x} e^{-i\vec p \cdot \vec x} \epsilon^{abc} \left[ \left[d^a(x)\right]^T (C\gamma_5 P_\pm) u^b(x) \right] d^c_\alpha(x),
\end{align}
where $C = \gamma_4 \gamma_2$, and $P_\pm = \frac{1}{2} (1 \pm \gamma_4)$ is the positive or negative parity projector. These single nucleon operators are then combined to form a $NN({}^3S_1)$ deuteron state, with quantum numbers $J^\pi = 1^+$ and $J_z = +\hbar$ with definite position at the source:
\begin{align}
    \mathscr{O}_{{}^3 S_1}(x)^\dagger &= \frac{1}{\sqrt{2}} \left( \left[ p^+(x)^\dagger \right]^T (C\gamma_3) n^+(x)^\dagger - \left[ n^+(x)^\dagger \right]^T (C\gamma_3) p^+(x)^\dagger \right), \label{eq:deu_src}
\end{align}
and definite momentum at the sink:
\begin{align}
    \mathscr{O}_{{}^3 S_1}(2 \vec p) &= \frac{1}{\sqrt{2}} \left( \left[ p^+(\vec p) \right]^T (C\gamma_3) n^+(\vec p) - \left[ n^+(\vec p) \right]^T (C\gamma_3) p^+(\vec p) \right).\label{eq:deu_snk}
\end{align}
In order to calculate the multi-nucleon correlation function, we use tensor e-graph optimised expressions as outlined in Ref.~\cite{humphrey_2022}.

\subsection{Feynman-Hellmann Theorem}

The direct calculation of the generalised Compton tensor defined in Eq.~(\ref{eq:structure_functions}) involves the computation of 4-point correlation functions, owing to the two non-local current insertions of the electromagnetic current $\mathcal{J}_\mu (z)$. An alternative approach developed in Refs.~\cite{utku_2020,utku_2022} is to apply the second-order Feynman-Hellmann Theorem to relate the 4-point function of interest to the second-order energy shift \cite{chambers_2017,agadjanov_2016}, subject to a perturbation to the fermion action:
\begin{align}
    S(\lambda) &= S_\textrm{unpert} + \lambda \int d^4 z \left( e^{i q \cdot z} + e^{-i q \cdot z} \right) \mathcal{J}_\mu(z),
\end{align}
where $\lambda$ is the coupling strength, and $\mathcal{J}_\mu(z) = Z_V \overline{q}(z) \gamma_\mu q(z)$. We begin by calculating the perturbed 2-point correlator:
\begin{align}
    C_\lambda(\vec p;t,q) &\coloneq \int d^3 \vec z e^{-i \vec p \cdot \vec z} \bra{\Omega_\lambda} A(\vec z; t) A(0)^\dagger \ket{\Omega_\lambda} \\
    &\simeq A_\lambda e^{-E_{A,\lambda} t}, \nonumber
\end{align}
with a selection of coupling strengths $\lambda$. By extracting each $E_{A,\lambda}$ using standard lattice QCD spectroscopic techniques, the second-order energy shift, $\frac{\partial^2 E_{A,\lambda}}{\partial \lambda^2}\Bigr|_{\lambda=0}$, can be extracted by a polynomial $\lambda$-fit. The second-order energy shift is then related to the Compton tensor via:
\begin{align}
    \frac{T_{\mu \mu}(p,q) + T_{\mu \mu}(p, -q)}{2} &= -E_A(\vec p) \frac{\partial^2 E_{A,\lambda}}{\partial \lambda^2}\Bigr|_{\lambda=0}, \label{eq:Tmumu}
\end{align}
where $E_A(\vec p)$ is the unperturbed energy of the state $A$. For the purposes of this work, $A(\vec z, t)$ is the $NN({}^3S_1)$ deuteron state defined in Eq. (\ref{eq:deu_src}).

\subsection{Extracting Moments of Compton Structure Functions}

Beginning with the Compton structure function decomposition of Eq.~(\ref{eq:structure_functions}), and making the assumptions that $q_3 = 0 = p_3$ and that the incoming and outgoing target is polarised in the z-direction, it is the case that:
\begin{align}
    T_{33}(p,q) &= \mathcal{F}_1(\omega, Q^2) + \frac{E_A^2 (q_1^2 + q_2^2)}{3} \frac{b_1(\omega,Q^2)}{(p \cdot q)^2}, \label{eq:T33} \\
    T_{00}(p,q) &= -\mathcal{F}_1(\omega, Q^2) + E_A^2 \frac{\mathcal{F}_2(\omega, Q^2)}{p \cdot q} - \frac{E_A^2 (q_1^2 + q_2^2)}{3} \frac{b_1(\omega,Q^2)}{(p \cdot q)^2}, \label{eq:T00}
\end{align}
noting that the 4-momentum vectors take the form $p^\mu = (E_A, \vec p) = (E_A, p_1, p_2, p_3)$ and $q^\mu = (0, \vec q) = (0, q_1, q_2, q_3)$. Clearly we can calculate $\mathcal{F}_2$ by rearranging Eqs.~(\ref{eq:T33}),~(\ref{eq:T00}) as:
\begin{align}
    \frac{\mathcal{F}_2(\omega, Q^2)}{\omega} &= \frac{Q^2}{2 E_A^2} \left[ T_{33}(p,q) + T_{00}(p,q) \right]. \label{eq:F2}
\end{align}
It is often convenient to consider the fixed $Q^2$ Taylor expansion of $\mathcal{F}_2(\omega,Q^2)$:
\begin{align}
    \mathcal{F}_2(\omega, Q^2) = \sum_{n=1}^\infty 4 \omega^{2n-1} M_{2n,A}^{(2)}(Q^2),
\end{align}
where the `moments' $M_{2n,A}^{(2)}$ for state $A$ are defined as:
\begin{align}
    M_{2n,A}^{(2)}(Q^2) &= \int_0^1 dx x^{2n-2} F_2(x,Q^2).
\end{align}
The lowest such moment is straightforwardly calculated by Eq.~(\ref{eq:F2}) as:
\begin{align}
    M_{2,A}^{(2)}(Q^2) &= \frac{Q^2}{8 E_A^2} \left[ T_{33}(p,q) + T_{00}(p,q) \right]\Bigr|_{\omega=0}. \label{eq:M2}
\end{align}
For the forthcoming results, we perform lattice QCD calculations in Euclidean time, so need to consider the Wick rotated form of Eq.~(\ref{eq:M2}). Under a Wick rotation $x_0 \to ix_4^{(E)}$, $T_{00} \to -T_{44}^{(E)}$ and $T_{33} \to T_{33}^{(E)}$, so that Eq.~(\ref{eq:M2}) becomes:
\begin{align}
    M_{2,A}^{(2,E)}(Q^2) &= \frac{Q^2}{8 E_A^2} \left[ T_{33}^{(E)}(p,q) - T_{44}^{(E)}(p,q) \right]\Bigr|_{\omega=0}. \label{eq:M2E}
\end{align}
For convenience, we henceforth drop the $(E)$ superscript and work entirely in Euclidean space coordinates.

\section{Results}

\begin{figure}[t]
    \centering
    \begin{subfigure}[t]{0.4\textwidth}
        \centering
        \includegraphics[width=1.2\linewidth]{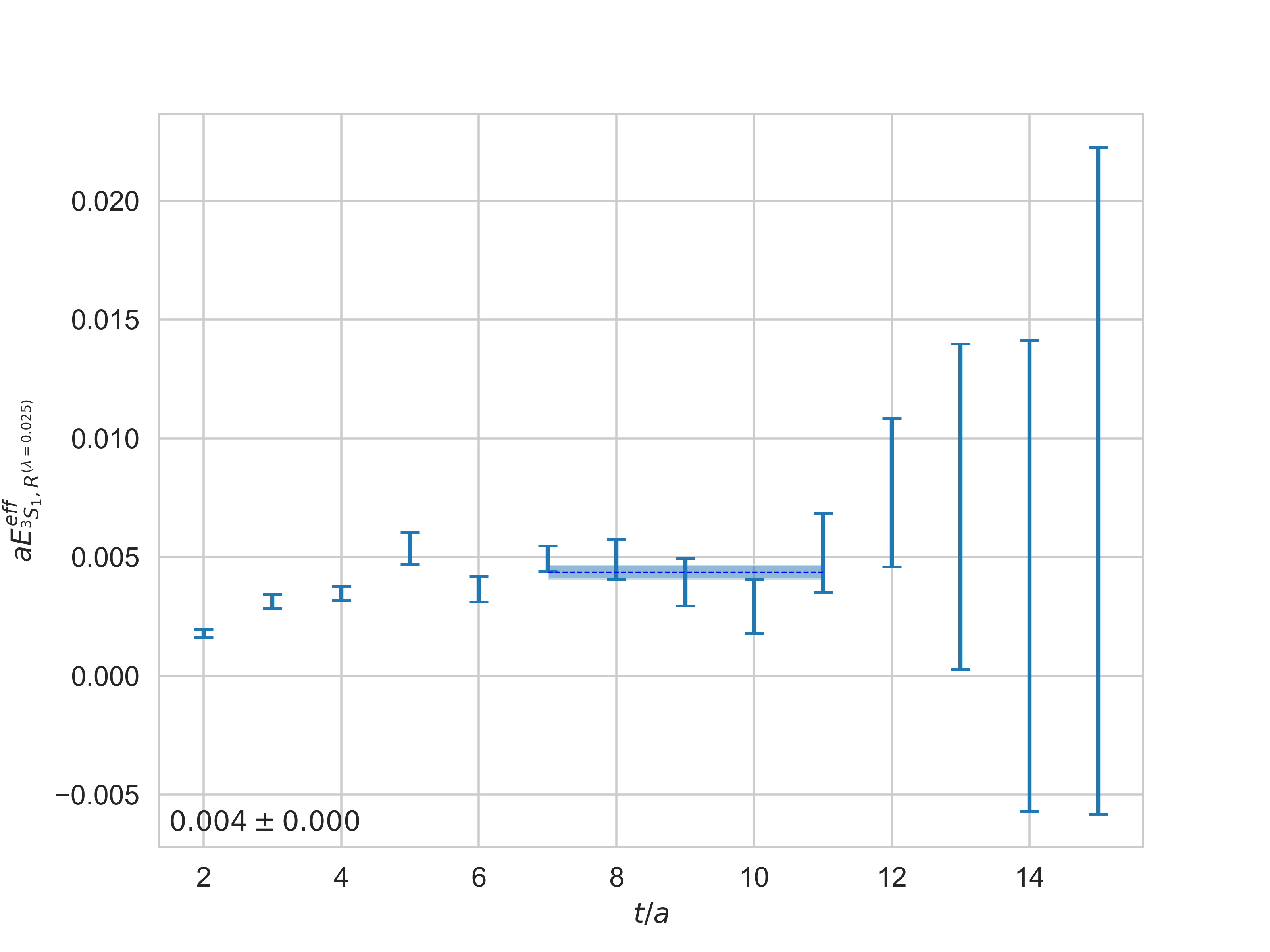}
        \caption{Measured $T_{44}$ ($T_{00} \to -T_{44}$ after Wick rotation) $NN({}^3S_1)$ effective energy shift (i.e. $\log\left( R^{(\lambda)}(t+1) / R^{(\lambda)}(t) \right)$ for $R^{(\lambda)}(t)$ as defined in Eq.~(\ref{eq:ratio}) for $\lambda = 0.025$.}
    \end{subfigure}%
    ~
    \begin{subfigure}[t]{0.1\textwidth}%
    ~
    \end{subfigure}%
    ~
    \begin{subfigure}[t]{0.4\textwidth}
        \centering
        \includegraphics[width=1.2\linewidth]{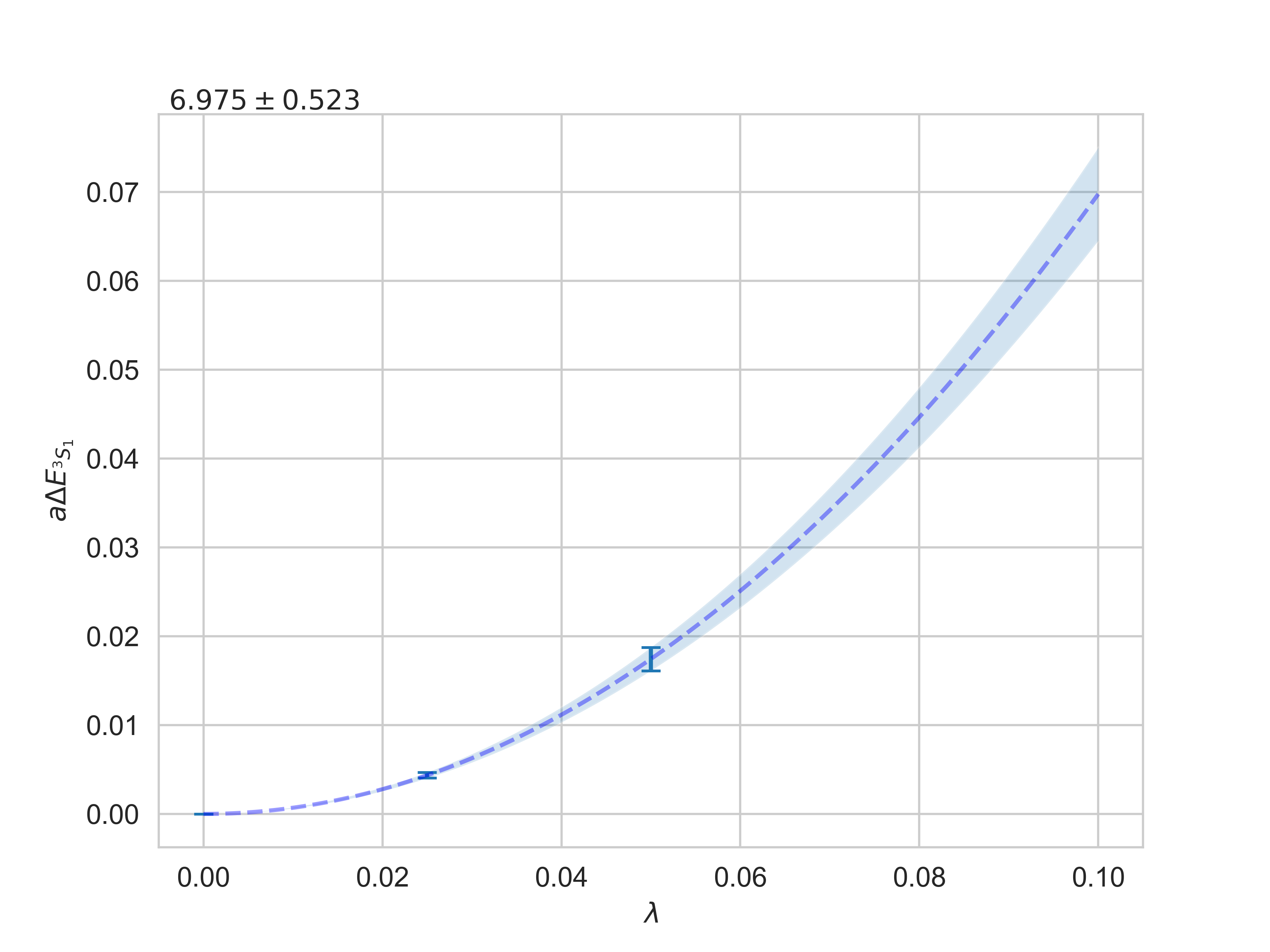}
        \caption{Quadratic $\lambda$ fit to $T_{44}$ $NN({}^3S_1)$ effective energy shifts, where measurements are taken at $\lambda = \pm 0.025, \pm 0.05$.}
    \end{subfigure}
    \begin{subfigure}[t]{0.4\textwidth}
        \centering
        \includegraphics[width=1.2\linewidth]{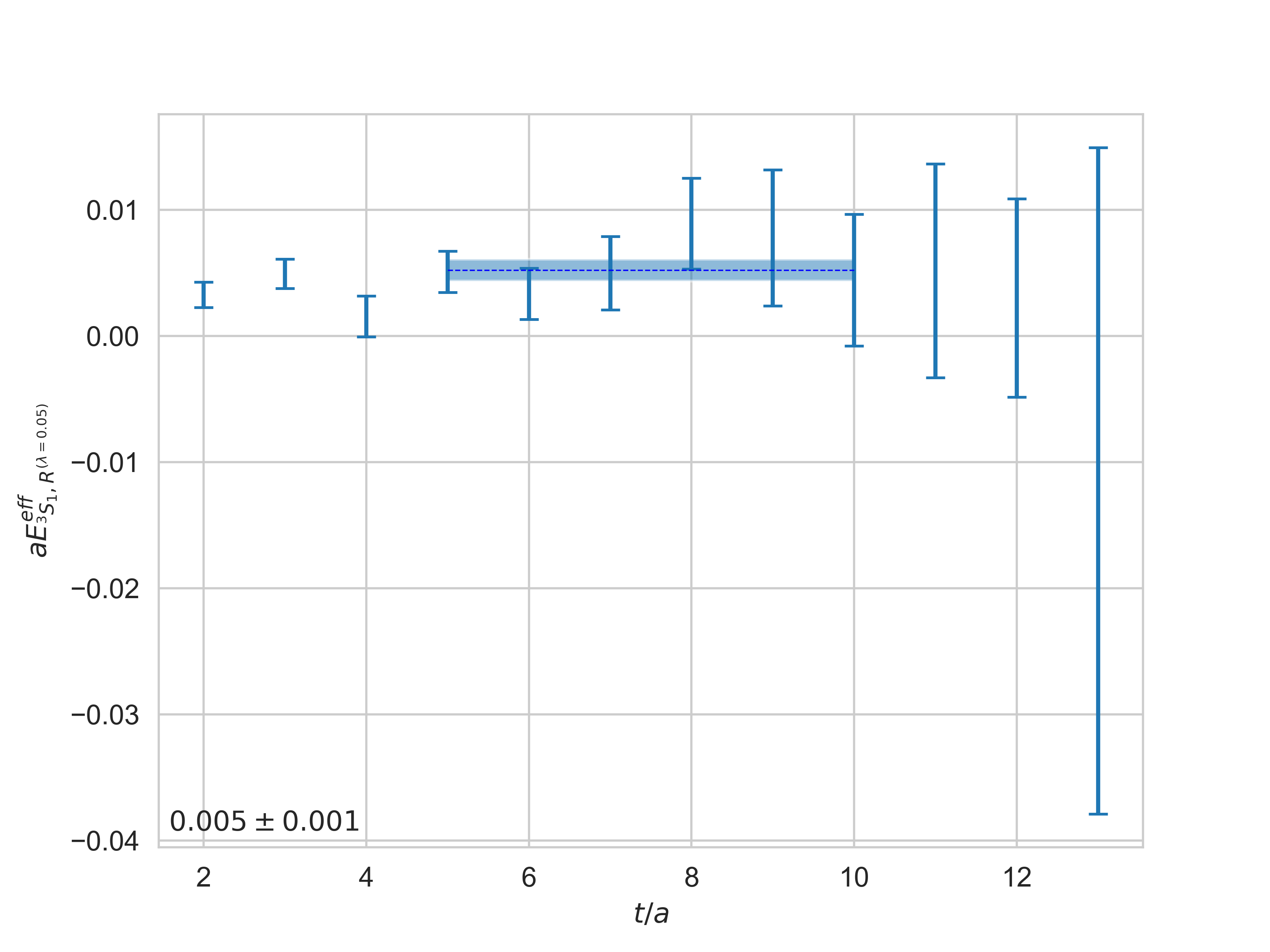}
        \caption{Measured $T_{33}$ $NN({}^3S_1)$ effective energy shift (i.e. $\log\left( R^{(\lambda)}(t+1) / R^{(\lambda)}(t) \right)$ for $R^{(\lambda)}(t)$ as defined in Eq.~(\ref{eq:ratio}) for $\lambda = 0.05$.}
    \end{subfigure}%
    ~
    \begin{subfigure}[t]{0.1\textwidth}%
    ~
    \end{subfigure}%
    ~
    \begin{subfigure}[t]{0.4\textwidth}
        \centering
        \includegraphics[width=1.2\linewidth]{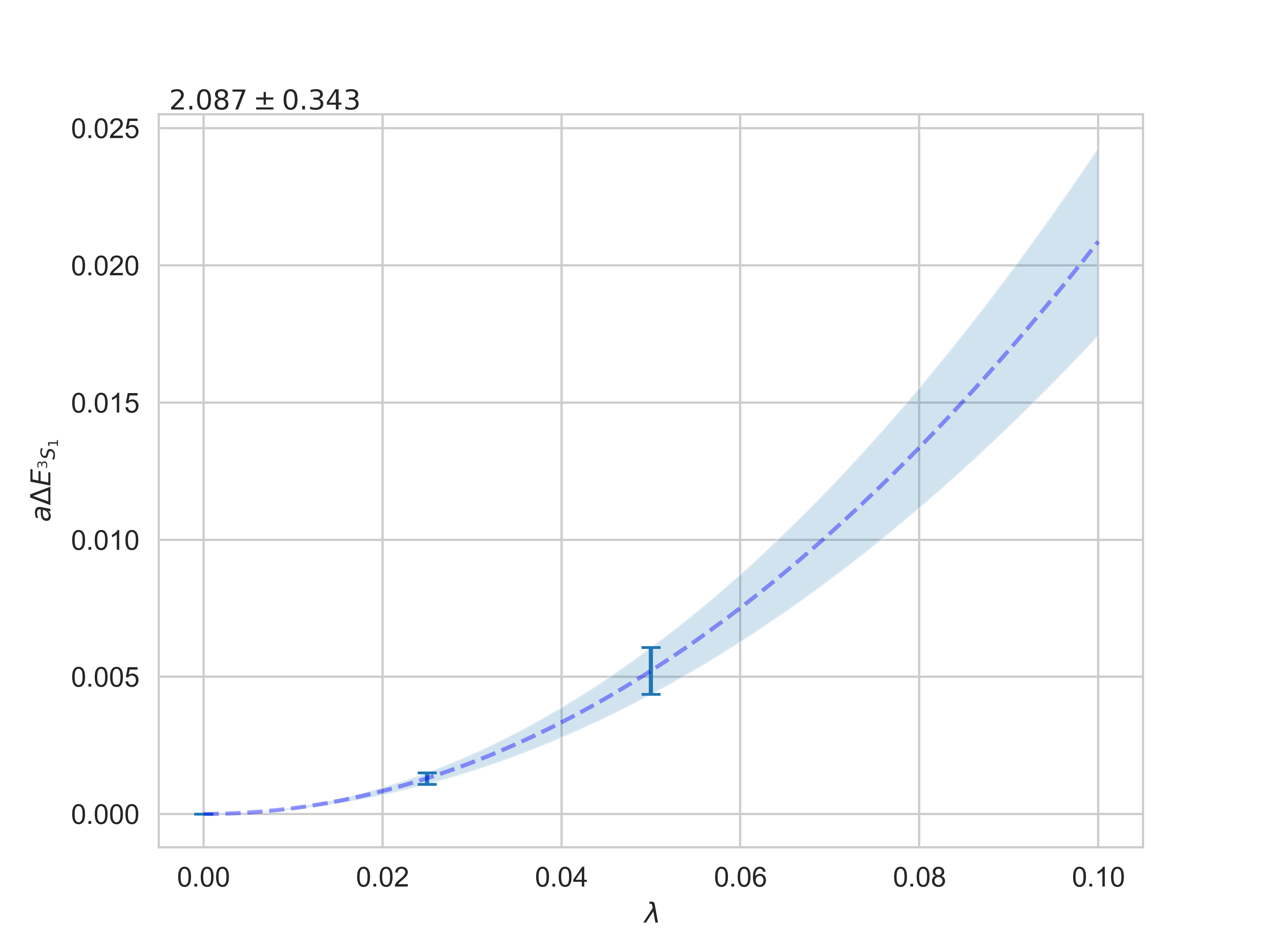}
        \caption{Quadratic $\lambda$ fit to $T_{33}$ $NN({}^3S_1)$ effective energy shifts, where measurements are taken at $\lambda = \pm 0.025, \pm 0.05$.}
    \end{subfigure}
    \caption{Effective energy shifts with extracted energy fit (first column), and corresponding quadratic $\lambda$ fits (second column).}
    \label{fig:fitting_tiles}
\end{figure}

We present calculations of $M_{2}^{(2)}$ for the $NN({}^3S_1)$ deuteron state defined in Eqs.~(\ref{eq:deu_src}), (\ref{eq:deu_snk}), by using the Feynman-Hellman lattice QCD procedure described in  §2.1, §2.2 to calculate $T_{\mu \mu}(\omega = 0)$ via Eq.~(\ref{eq:Tmumu}) with coupling strengths $\lambda = \pm 0.025, \pm 0.05$, and kinematics $\vec p = 0$, $\vec q = (4,1,0) 2\pi/L$. In order to extract the perturbed energy shift from two-point functions $C_\lambda(t)$, we calculate the ratio:
\begin{align}
    R^{(\lambda)}(t) &= \frac{\langle C_\lambda(t) \rangle \langle C_{-\lambda}(t) \rangle}{\left[\langle C_{\lambda = 0}(t) \rangle \right]^2}, \label{eq:ratio} \\
                    &\sim A_\lambda e^{-2 \Delta E_{A,\lambda} t}.
\end{align}
A plateau in the effective energy (i.e. $\log\left( R^{(\lambda)}(t) / R^{(\lambda)}(t+1) \right)$) corresponds to the perturbed energy shift $2 \Delta E_{A,\lambda}$. Fig.~\ref{fig:fitting_tiles} depicts the process of extracting $T_{44}$, $T_{33}$ for $NN({}^3S_1)$ by first extracting the effective energy plateau for each $R^{(\lambda)}$, and then fitting $\frac{\partial^2 E_{A,\lambda}}{\partial \lambda^2}\Bigr|_{\lambda=0}$ via a quadratic $\lambda$ fit. \par

\begin{figure}[H]
    \centering
    \includegraphics[width=0.9\textwidth]{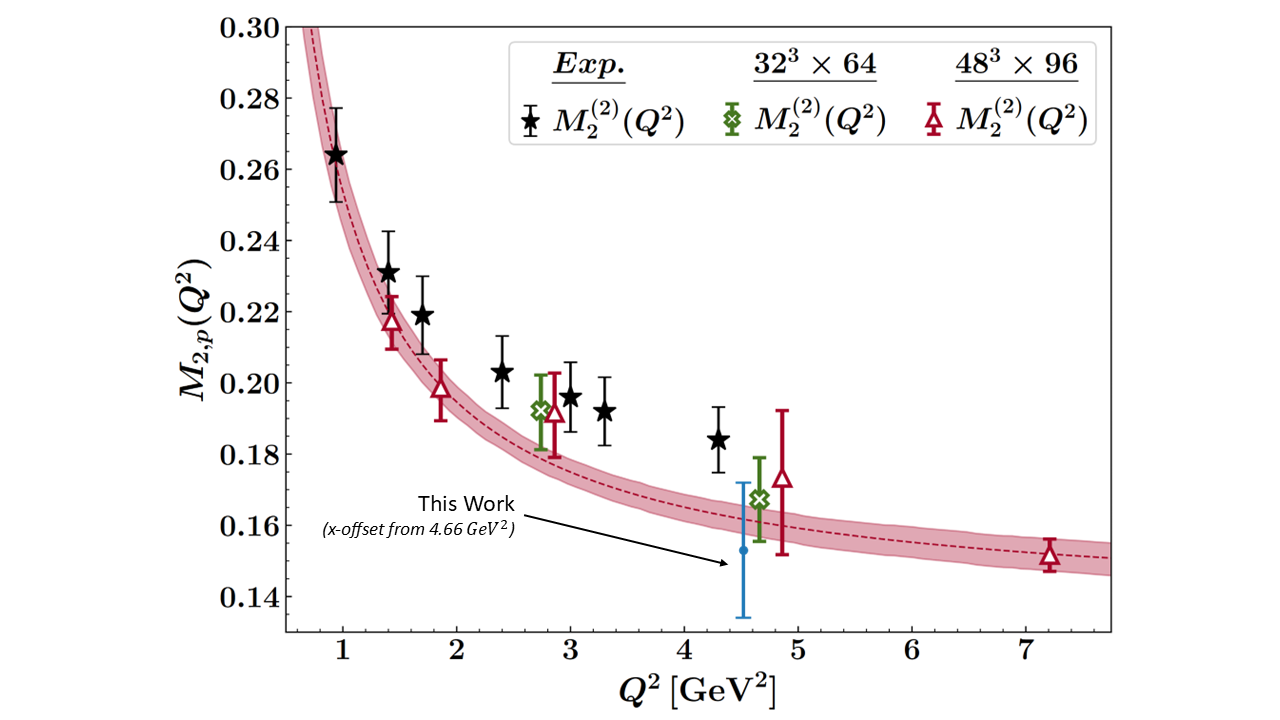}
    \caption{Deuteron $M_2^{(2)}$ result in blue superimposed on reproduction of Figure 2 of Ref.~\cite{utku_2022}. $Q^2$ dependence of the lowest moment of $\mathcal{F}_2$ for the proton \cite{utku_2022}. Filled stars are the experimental Cornwall-Norton moments of $\mathcal{F}_2$ taken from Table I of Ref.~\cite{armstrong_2001}. We have assigned a 5\% error to the experimental moments as indicated in Ref.~\cite{armstrong_2001}. The red band is the $1/Q^2$ fit to the $48^3 \times 96$ data points.}
    \label{fig:deu_M_fig}
\end{figure}

From the quadratic fits in $\lambda$, we calculate $M_{2,NN({}^3S_1)}^{(2)}(Q^2=4.66\ GeV^2) = 0.153 \pm 0.019$ for our $NN({}^3S_1)$ deuteron state using Eq.~(\ref{eq:M2E}). Fig.~\ref{fig:deu_M_fig} shows this $NN({}^3S_1)$ value superimposed on a reproduction of Figure 2 from Ref.~\cite{utku_2022}, which depicts lattice QCD proton calculations of $M_{2,p}^{(2)}$ (the lowest moment of $\mathcal{F}_2$) for a range of $Q^2$ values and two lattice volumes ($32^3 \times 64$ with $a=0.074(2)$ fm and $48^3 \times 96$ with $a=0.068(3)$ fm). Note in particular that at this level of statistics and choice of $Q^2$, $M_{2}^{(2)}$ for our deuteron is indistinguishable from the proton result.

Fig.~\ref{fig:deu_M_fig} shows that there is reasonable agreement between experimental and lattice QCD results despite the relatively heavy $470$ MeV pion mass of the lattice QCD simulations. Note that there are hints both in the experimental and lattice QCD data of non-trivial $Q^2$ dependence in the region $Q^2 \in [2,5]$ Ge$V^2$. The $Q^2 \to \infty$ limit is well understood by the PDF moment. The aim of our programme of work is to produce an equivalent plot to Fig.~\ref{fig:deu_M_fig} of the $Q^2$ dependence of $M_{2,A}^{(2)}$ for the $NN({}^3S_1)$ deuteron state defined in Eqs.~(\ref{eq:deu_src}),~(\ref{eq:deu_snk}), and compare to the single nucleon results.

\section{Conclusion}
We presented a calculation of the lowest moment of the Compton structure function $\mathcal{F}_2$ for a multi-nucleon $NN({}^3S_1)$ deuteron state using Feynman-Hellmann lattice QCD techniques, and compared this result against previous proton calculations, showing a consistent value. Future work would aim to increase the number of measurements in order to resolve the detailed structure of $\mathcal{F}_2$ as a function of $Q^2$ and number of nucleons $A$.



\section{Acknowledgments}
The numerical configuration generation (using the BQCD lattice QCD program \cite{haar_2017}) and data analysis (using the Chroma software library) was carried out on the Extreme Scaling Service (DiRAC, EPCC, Edinburgh, UK), the Data Intensive Service (DiRAC, CSD3, Cambridge, UK), the Gauss Centre for Supercomputing (NIC, J\"{u}lich, Germany), the NHR Alliance (Germany), and resources provided by the NCI National Facility in Canberra, Australia (supported by the Australian Commonwealth Government), the Pawsey Supercomputing Centre (supported by the Australian Commonwealth Government and the Government of Western Australia), and the Phoenix HPC service (University of Adelaide). NH is supported by an Australian Government Research Training Program (RTP) Scholarship. RH is supported by STFC through grant ST/X000494/1. PELR is supported in part by the STFC under contract ST/G00062X/1. GS is supported by DFG Grant SCHI 179/8-1. KUC, RDY and JMZ are supported by the Australian Research Council grants DP190100297, DP220103098, and DP240102839. We would like to thank Will Detmold for valuable discussions.

\bibliographystyle{JHEP}
\bibliography{refs}



\end{document}